\documentclass[seceq,epsf]{ptptex}
\usepackage{wrapft}

\usepackage{graphicx}
\def\undertext#1{$\underline{\hbox{#1}}$}
\def\gtsim{\mathrel{\hbox{\raise0.2ex
\hbox{$>$}\kern-0.75em\raise-0.9ex\hbox{$\sim$}}}}
\def\ltsim{\mathrel{\hbox{\raise0.2ex
\hbox{$<$}\kern-0.75em\raise-0.9ex\hbox{$\sim$}}}}
\def\leqqq{\mathrel{\hbox{\raise0.2ex
\hbox{$<$}\kern-0.75em\raise-0.9ex\hbox{$=$}}}}
\def\geqqq{\mathrel{\hbox{\raise0.2ex
\hbox{$>$}\kern-0.75em\raise-0.9ex\hbox{$=$}}}}
\def \Romannumeral(#1) {\uppercase\expandafter{\romannumeral#1}}

\def \Romannumeral(#1) {\uppercase\expandafter{\romannumeral#1}}
\def \Rn(#1) {\uppercase\expandafter{\romannumeral#1}}
\def\Fig(#1){$${\overline{\underline{\rm Fig.{\ \ #1}}}}$$} %%%%%%Fig%%
\def\Tab(#1){$${\overline{\underline{
 \rm Table\ \  \uppercase\expandafter{\romannumeral#1}}}}$$} %%%%%%Table%%
\def \a{\alpha}
\def \b{\beta}

\def \t{\tau}

\def \m{ \mu }
\def \n{ \nu }

\def \t{\theta}

\def \fn1{N_{f_1}}
\def \fn2 {N_{f_2}}

\def \Romannumeral(#1) {\uppercase\expandafter{\romannumeral#1}}
\def \Rn(#1) {\uppercase\expandafter{\romannumeral#1}}

\def \mib #1 {$\mbox{\boldmath $ #1 $}$}
\def\sqr#1#2{{\vcenter{\vbox{\hrule height.#2pt \hbox{\vrule width.#2pt 
height#1pt \kern#1pt \vrule width.#2pt}\hrule height.#2pt}}}}

\title{%        %You can use \\ for explicit line-break
The $\theta$-term,  CP$^{N-1}$ Model and the Inversion Approach\\
 in the Imaginary $\theta$ Method %
}

\author{%       %Use \scshape  for the family name
Masahiro \textsc{IMACHI}$^{1,}$
\footnote{E-mail:masaimac@tkg.bbiq.jp} 
\\ 
Hitoshi \textsc{KAMBAYASHI}$^{2}$
\\ 
Yasuhiko \textsc{SHINNO}$^{3,}$
\footnote{E-mail:shinno@dirac.phys.saga-u.ac.jp} 
and
Hiroshi \textsc{YONEYAMA}$^{3,}$
\footnote{E-mail:yoneyama@cc.saga-u.ac.jp} 
\\ 
}

\inst{%         %Affiliation, neglected when [addenda] or [errata]
$^1$Department of Physics, Yamagata University, Yamagata 990-8560, Japan
\\
$^{2}$ Towa System Inc., Tokyo 101-0052, Japan
\\
$^{3}$Department of Physics, Saga University, Saga 840-8502, Japan
}

\abst{%         %this abstract is neglected when [addenda] or [errata]
The weak coupling region of CP$^{N-1}$ lattice field theory with the $\theta$-term is 
investigated. Both the usual real theta method and the imaginary theta method are studied. The latter was first proposed by Bhanot and David. Azcoiti et al.~proposed an inversion approach based on the imaginary theta method.  The role of the inversion approach is investigated in this paper. A wide range of values of $h=-{\rm Im} \theta$ is studied, where $\theta $ denotes the magnitude of the topological term. Step-like behavior in the $x$-$h$ relation (where $x=Q/V$, $Q$ is the topological charge, and $V$ is the two dimensional volume)  is found 
in the weak coupling region. The physical meaning of the position of the step-like behavior is discussed. The inversion approach is applied to weak coupling regions. \par

}

\begin{document}

\maketitle
%\input{added}

%%%%%%%%%%%%%%%%%%%%%%%%%%%%%%%%%%%%%%%%%%%%%%%%%%%%%%%%%%%
\section{Introduction}

%%%%%%%%%%%%%%%%%%%%%%%%%%%%%%%%%%%%%%%%%%%%%%%%%%%%%%%%%%%%%%%%%%%%%
%\subsection{Introduction}
%{\gt Introduction}\par
The two-dimensional lattice CP$^{N-1}$ model with a $\theta$-term is investigated. 
The problem of obtaining the partition function $Z(\theta)$ numerically stems from the 
difficulty in treating the  complex valued Boltzmann weight. This difficulty is 
avoided by expressing  $Z(\theta)$  as a Fourier series,
\cite{rf:Wiese}\par
$$Z(\theta)=\sum_Q P(Q) e^{i \theta Q},$$
where $P(Q) $ is the topological charge distribution, i.e., 
the probability of 
finding a topological charge $Q$ in the system at $\theta=0$. \par
In the strong coupling region, $P(Q) $ can be  approximately expressed as a Gaussian function, 
$P(Q)\propto \exp (-\frac{\alpha}{V} Q^2) $, and a first-order phase
 transition 
at $\theta= \pi $ is obtained.\cite{rf:Wiese}\tocite{rf:PS} In the weak coupling region, $P(Q) $ exhibits  behavior that differs greately from the Gaussian form. Instead of a quadratic $Q$  dependence, an almost linear form is  
 found\cite{rf:IKY}\footnote{ See Eq. (4.9) of Ref. \citen{rf:IKY}.}   in the exponent of $P(Q)$: 
\begin{eqnarray}
P(Q) \propto c^{|Q|}=e^{|Q| \ln c}.
\label{eq:Pqlinear } 
\end{eqnarray}%
 In the weak coupling region, $c$  was found to be a quite small constant. This linear exponent is a simplified typical form. \par
Bhanot and David  first proposed the imaginary theta method\cite{rf:BhDa}, in which the $\theta$ parameter 
is taken to be purely imaginary, 
$$\theta =-i h,$$  
with $h$ being a real parameter. Azcoiti et al. proposed an  inversion approach\cite{rf:ACGL} based on the imaginary theta method.\par
In the imaginary $\theta$ case, numerical simulations can be performed for 
$Z(h)$, since the Boltzmann factor becomes real in this  case, and we have 

$$ Z(h)=\int {\cal D}z {\cal D}{z^*}  e^{-S(z,{z^*})+h Q(z,{z^*}) },$$ 
where $S$ is an action and $z \ ( z^*)$ denotes appropriate fields (their complex conjugates).
But it seems that the meaning of the inversion approach\cite{rf:ACGL} 
 is not   well understood. 
 For this reason, the role of the inversion approach\cite{rf:ACGL} based on the imaginary theta method\cite{rf:BhDa} is  investigated in this paper.\par
We  performed a  numerical analysis with $\theta =-i h$ for both 
strong and weak coupling regions. After presenting the results of this analysis, we  discuss  the 
 meaning of the approach used by Azcoiti et al.  Our understanding of the 
 meaning  of their approach is summarized as follows.\par

\begin{itemize}
%\begin{enumerate}
\item[1.] In some cases, real theta results can be obtained from imaginary theta results by analytic continuation. However, this  is not true in other cases.  What,  then, does the imaginary theta method mean? It does not mean analytic continuation at nonzero theta. The inversion approach is nothing but  one of the 
fitting methods   of the 
  topological charge
 distribution $P(Q)$  \undertext { at $\theta=0$ }. This is shown in \S 3.\par
\item[2.]The imaginary theta method is suited to determining the  $h$-dependence for a wide range of values of $h$. 
In particular, the  $x$-$h$ relation for a wide range of values of $h$, and thus a wide 
range of  values of $x$,  is obtained, 
where $x=Q/V$. From this $x(h)$-$h$ relation, the inversion approach leads to the $h(x)$-$x$ relation. \par
\item[3.]In the strong coupling region, the Gaussian form of $P(Q)$ is 
reconfirmed using the inversion approach.\par 
\item[4.]In the weak coupling region, we have found ``step-like behavior"
 in  the $x$-$h$
 relation. The position of a step gives  the value of the parameter  $c(\beta)$, 
 where $\beta$ represents the inverse coupling constant of the CP$^{N-1}$ model.  
 The parameter $c(\beta)$ is that found in our previous analysis of the
 topological charge distribution\cite{rf:IKY} at $\theta=0$, $P(Q) \propto 
 c(\beta)^{|Q|}$.\par
\item[5.]In the weak coupling region, fluctuation of the field variables $z$ and
 ${z^*}$ is greately suppressed,  and thus the probability of topological charge 
 excitation is also greately suppressed in  regions of small $h$. The parameter $c(\b)$ is a measure of the degree of
 suppression of the topological charge excitation.\par
 \item[6.]The  $h(x)$-$x$ relation is obtained from the $x(h)$-$h$ relation in the inversion approach. The functional form of $h(x)$ is obtained by fitting to the data with the appropriate function. 
  Actually, at the end of \S\ref{sec:numerical}, $f(x)$ in the weak coupling region is obtained by integrating $h(x)=df(x)/dx$  in the inversion approach [see Eq.~(\ref{E})], and  the leading term in that result reproduces the result of the linear model [exponent of Eq.~(\ref{eq:Pqlinear }) ]. 
 
 \par
%\item{()}
%\item{()}
\end{itemize}
%\end{enumerate}
\vskip .5cm

This paper is
organized as follows. The inversion approach based on the imaginary theta method is explained in 
\S 2.
 The results of the numerical calculation are presented in \S 3. Conclusions and discussion  are given in \S 4.

%%%%%%%%%%%%%%%%%%%%%%%%%%%%%%%%%%%%%%%%%%%%%%%%%%%%%%%%%%%
\section{Inversion approach in the imaginary theta method}
\subsection{ Formulation} 
%%%%%%%%%%%%%%%%%%%%%%%%%%%%%%%%%%%%%%%%%%%%%%%%%%%%%%%%%%%%%%%%%%%%%
Much progress has been made in non-Abelian lattice gauge theory; e.g., asymptotic
 freedom was confirmed by the observation of string tension.\cite{rf:CR}  It appears possible that quark confinement can be  explained  as an ``area law"  in the lattice formulation. Instanton excitation allows the   existence of the topological term, namely a theta term in the action. However, lattice field theory  with  the theta term is not well understood, 
  because   the Euclidean formulation introduces a complex Boltzmann factor, and it 
  does not allow  direct Monte Carlo simulation.
Bhanot and David  first  introduced a purely imaginary theta parameter and studied the $O$(3) non-linear sigma model\cite{rf:BhDa}. 
 If we take $\theta$ to be purely imaginary, the Boltzmann weight becomes a real 
 positive quantity, and this allows dirfect numerical simulation. 
 Azcoiti et al. introduced the inversion approach based on the imaginary theta method in Ref. \citen{rf:ACGL}. 
However it seems that the meaning of the inversion approach  is not  well  understood.
   In order to understand the meaning of the inversion approach,  we employ 
   the imaginary theta method in both the strong  and the weak 
   coupling regions and study the inversion approach. We  then compare the imaginary  and real theta  methods and study the role of the inversion approach. \par  
We  begin with the real $\theta$ case. 
%%%mod0925 from
 The  CP$^{N-1}$  model with the $\theta$-term on a two-dimensional Euclidean lattice is
   considered.\cite{rf:HITY}\cite{rf:Seiberg} The action with the $\theta$-term is defined by
\begin{eqnarray}
  S_{\theta}(z, z^*)=S(z, z^*)-i \theta Q(z, z^*),
\label{2.1} 
\end{eqnarray}%
 where 
\begin{eqnarray}
S(z, z^*)=\b \sum _{n, \m}\left \{1-\sum_{\a=1}^N |z^*_\a (n) z_\a (n+{\hat\m}) |^2 
  \right \}
\label{2.2} 
\end{eqnarray}%
  is the action and $z_{\a}(n)( \a=1,\cdots N)$ denotes a CP$^{N-1}$ field 
 on each site $n$. The site $n+{\hat\m}$ is the site nearest to $n$ in the direction $\m$. 
 The topological charge $Q(z, z^*)$ is defined as
\begin{eqnarray}
Q(z, z^*)= \frac{1}{2\pi}\sum_{\Box}A_{\Box},
\label{2.3} 
\end{eqnarray}%
\begin{eqnarray}
   A_{\Box}(z, z^*)  =\frac{1}{2}\sum_{\m, \n}\{A_{\m}(n)+ A_{\n}(n+{\hat\m})- A_{\m}(n+{\hat\n})
   - A_{\n}(n)
   \}\epsilon_{\m\n}, 
 \label{2.4} 
\end{eqnarray}%
where the quantities $A_{\m}(n)$ are defined as 
\begin{eqnarray}
\exp (i A_{\m}(n))={ z}^\dagger(n)z(n+{\hat\m})/|{ z}^\dagger(n) z(n+{\hat\m})|
\label{} 
\end{eqnarray}%
by the CP$^{N-1}$ field $z$. For the CP$^{N-1}$ model, the complex field $z_{\a}(n)$ satisfies the equation
\begin{eqnarray}
{ z}^\dagger(n)z(n)=\sum^{N}_{\a=1} { z}^*_{\a}(n)z_{\a}(n)=1.
\label{} 
\end{eqnarray}%
%

%%%mod0925  to here

The partition function for the  two-dimensional 
CP$^{N-1}$ field theory is given by
\begin{eqnarray}
% 
%\[
\begin{array}{ll}
Z_V (\theta)&=\frac{\displaystyle \int {\cal D}z {\cal D}{z^*}  
\exp (-S(z, {z^*})+ i \theta Q(z, {z^*}) )}{\displaystyle 
\int {\cal D}z {\cal D}{z^*}  \exp (-S(z, {z^*}))} \\ 
 &=\displaystyle \sum _{Q= {\rm integer}} \frac{\displaystyle \int
  ({\cal D}z {\cal D}{z^*}) ^{(Q)}\exp (- S(z, {z^*})+ i \theta  Q(z, {z^*})) }
  {\displaystyle \int {\cal D}z {\cal D}{z^*}  \exp (- S(z, {z^*})) } \\
 &=\displaystyle \sum _{Q= {\rm integer}} P (Q) e^{i \theta Q},
\end{array}
%\]
\label{2.7} 
\end{eqnarray}%
where $({\cal D}z {\cal D}{z^*})^{(Q)}$ denotes the constrained measure in which $ Q(z, {z^*})=Q$.
Here, $P (Q)$ is the topological charge distribution estimated with the action for $\theta=0$ as

\begin{eqnarray}
P (Q)=\frac{\displaystyle \int
  ({\cal D}z {\cal D}{z^*}) ^{(Q)}\exp (-S) }
  {\displaystyle \int {\cal D}z {\cal D}{z^*}  \exp (-S) }.
\label{2.8} 
\end{eqnarray}%
It  satisfies the relation
\begin{eqnarray}
\displaystyle \sum_{Q} P (Q)=1.
\label{2.9} 
\end{eqnarray}%
Once the  topological charge distribution is known, the partition function at any 
$\theta$ can be obtained from  the Fourier series  for the case of real $\theta$:
\begin{eqnarray}
Z_V(\theta)=\displaystyle{ \sum_{Q} P (Q) e^{i\theta Q}}
= \displaystyle{\sum_{x_Q} \exp(-Vf_V(x_Q))e^{i \theta V x_Q}},
\label{2.10} 
\end{eqnarray}%
where 
\begin{eqnarray}
x_Q=Q/V, \qquad f_V(x_Q)=-\frac{1}{V}\ln P(Q).
\label{2.5} 
\end{eqnarray}%
Now we  introduce an imaginary $\theta$\cite{rf:ACGL}. Setting  
$\theta=-i h\  (with h$ real), we have  
\begin{eqnarray}
\begin{array}{ll}
Z_V(h)&=\displaystyle {\frac{\int {\cal D}z {\cal D}{z^*} \exp (-S(z, {z^*})+h  Qz, {z^*}))}{\int {\cal D}z {\cal D}{z^*} \exp (-S) }}\cr
&=\displaystyle \sum_{{ x}_Q}\exp(-Vf_V({ x}_Q)) 
e^{h V { x}_Q}.
\label{2.12} 
\end{array}
\end{eqnarray}%
%%%mod 925 from
Thus  $h$ plays the role of the external source for the topological charge $x_Q$. 
Once a constant background field $h$ is given, the expectation value ${\bar x}(h)$ of the 
topological charge per unit volume is given by
\begin{eqnarray}
\begin{array}{ll}
{\bar x}(h)  = \frac{\displaystyle \int {\cal D}z {\cal D}{z^*} \frac{ Q(z, {z^*})}{V} \exp (-S_h)}{\displaystyle \int {\cal D}z {\cal D}{z^*}
 \exp (-S_h)}
=\displaystyle \frac{\sum_{x_Q} x_Q P_h(Q)}{\sum_{x_Q}  P_h(Q)},
\label{} 
\end{array}
\end{eqnarray}%
where
\begin{eqnarray}
P_h(Q)=\int ({\cal D}z {\cal D}{z^*}) ^{(Q)}\exp (-S_h)/\int {\cal D}z {\cal D}{z^*} \exp (-S),
\label{} 
\end{eqnarray}%
and
\begin{eqnarray}
S_h(z, {z^*})=S(z, {z^*})-h  Q(z, {z^*}).
\label{} 
\end{eqnarray}%

From this form, we can conclude that the  imaginary theta method is a kind of ``trial function"
(subtraction) method, in which the action is replaced by that with the subtraction term $S_{\rm trial}$  
$$ S_{\rm eff}=S_h=S-S_{\rm trial}.$$
The ``trial (subtraction) function"  is 
taken as a special form in the imaginary theta method, \par

$$ S_{\rm trial}(z, {z^*})=-h  Q(z, {z^*}).$$

%%%mod 925 to here

Now we  explain the inversion approach. 
When the volume $V$ is large, $x_Q$ is almost continuous, and $x_Q$ in the sum in Eq. (\ref{2.12}) can be  
approximated  by  ${\bar x}_Q$, the value of  $x_Q$ at which   
$\exp\{ -V(f_V (x_Q)-h x_Q)\}$ is maximal. This  saddle point method 
gives  
\begin{eqnarray}
Z_V(h)   \propto \displaystyle \exp(-V(f_V({\bar x}_Q)- h {\bar x}_Q )),
\label{zvh} 
\end{eqnarray}%
with 
\begin{eqnarray}
\frac{d g(x_Q)}{dx_Q} =0 \quad  {\rm at} \quad x_Q={\bar x}_Q,
\label{2.18} 
\end{eqnarray}%
where 
\begin{eqnarray}
g(x_Q)=f_V (x_Q)- h x_Q.
\label{2.19} 
\end{eqnarray}%
Equation (\ref{2.18}) gives 
\begin{eqnarray}
\left[ \frac{df_V (x_Q)}{dx_Q} \right ] _{x_Q={\bar x}_Q}=h.
\label{dfdxQ} 
\end{eqnarray}%
The quantity ${\bar x}_Q$ is  the expectation value of the topological charge 
(per unit volume).
\begin{itemize}
\item[1)]The expectation value ${\bar x}_Q$  for a 
given background $h$ can be obtained by numerical simulation. Specifically, ${\bar x}$ is regarded as  a function of $h$, and the ${\bar x(h)}$-$h$ relation can be  obtained.\par
\item[2)] On one hand, due to Eq.~(\ref{dfdxQ}), $h$ is the first derivative of $f_V (x)$ at ${\bar x}_Q$. Hereafter, the suffix $V$ is omitted.\par
\item[3)]We plot an illustrative example of ${\bar x}$ as a function of $h$  in Fig.~1.
 Exchanging $h$ and  $x$, we obtain Fig.~2. This is the ``inversion" of the ${\bar x(h)}$-$h$ relation to the $h(x)$-$x$ relation. In other words, $h$ is now a function of $x$.
\par
\end{itemize}

%%%%%%%%%%%%%%%%%%
\input epsf.tex
       \begin{figure}[htb]
            \parbox{\halftext}{%   %\def\halftext{.471\textwidth}
               \epsfysize= 6 cm
               \hskip.05cm
                 %\epsfbox{Spr1.eps}
                  \epsfbox{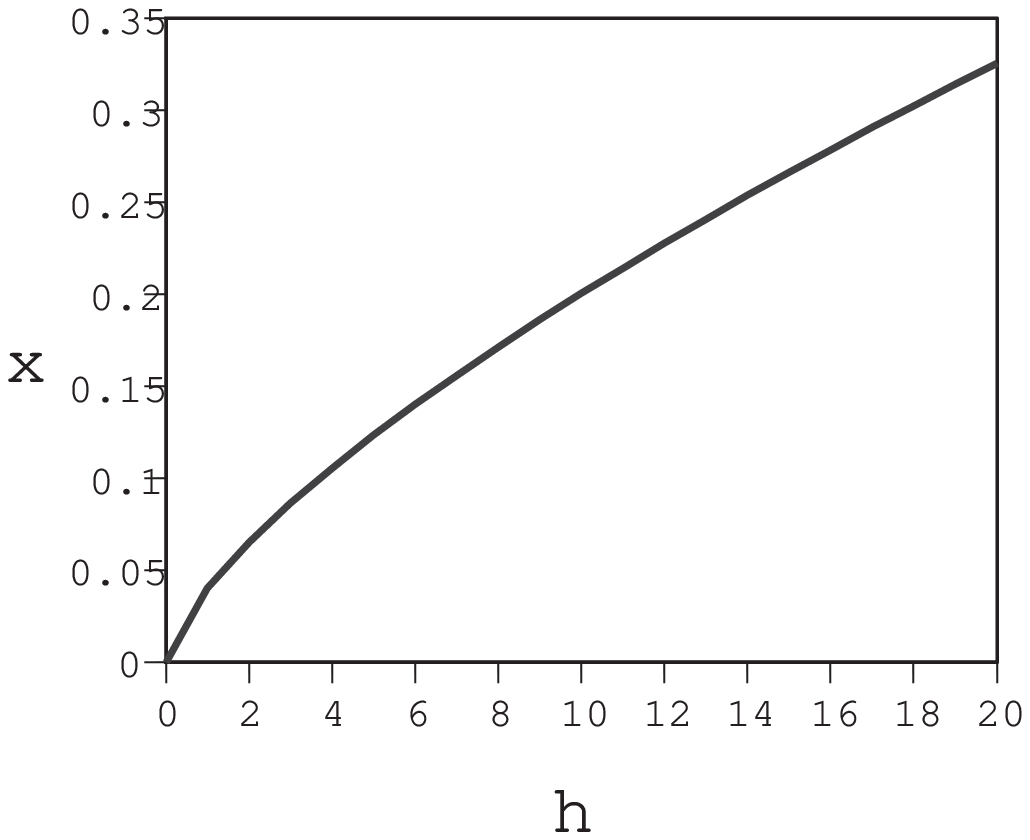}
                     %  \figurebox{6cm}{2cm}
             \label{fig1}
               \caption{An illustration of the expectation value $x$ as a function of the background field $h$.}}
            \hfill
            \parbox{\halftext}{
               \epsfysize= 6 cm
               \hskip.05cm
                 %\epsfbox{Spr1.eps}
                  \epsfbox{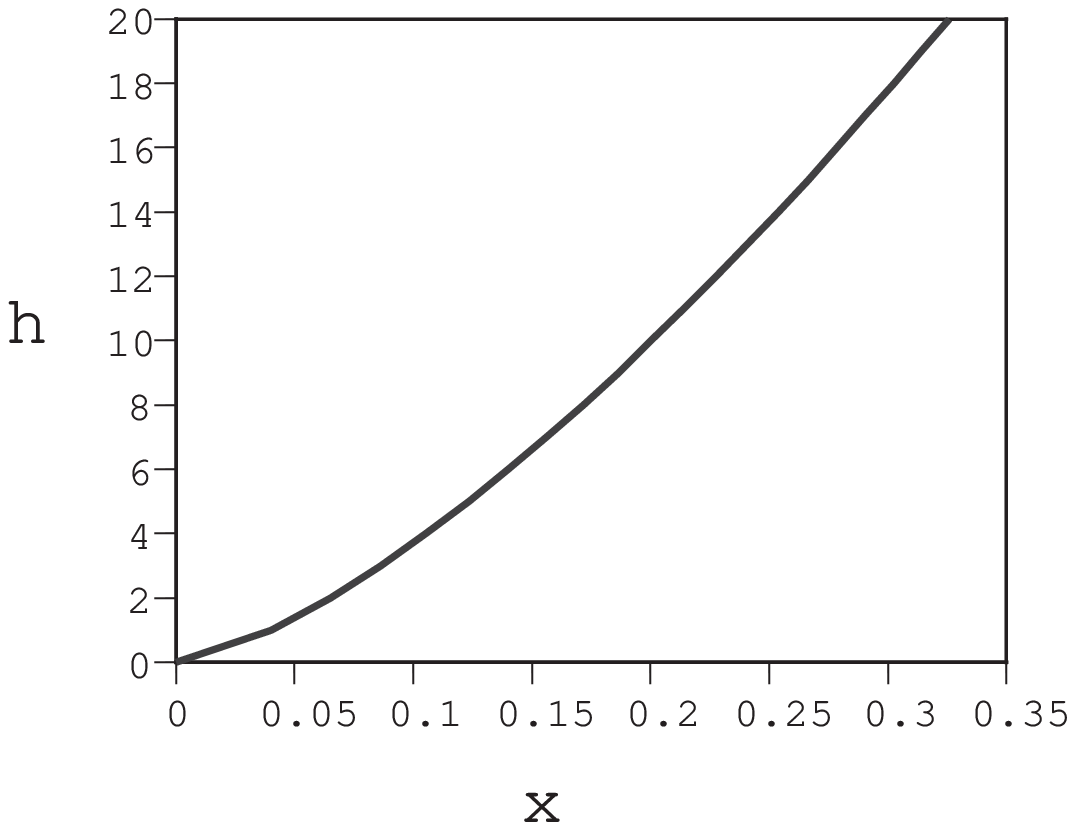}
                     %  \figurebox{6cm}{2cm}
               % \figurebox{6cm}{2cm}
                \caption{Exchanging $x$ and $h$ in Fig.1, we obtain $h=f'(x)$ as a function of $x$.}}
            \label{fig2}
           \end{figure}

%-------------------------------------------------------------------------
\vskip 0.5cm
 From the relation 
\begin{eqnarray}
 h=\left. \displaystyle\frac{df(x)}{dx} \right |_{x=\bar x},
 \label{dfdx}
 \end{eqnarray}
we have $df(x)/dx$ as a function of $x$. That is, fitting $h$ by an appropriate function of ${\bar x}$, we obtain the functional form of $df(x)/dx$. By integrating this over $x$,
\begin{eqnarray}
\int^{x}_{0} \frac{df(x')}{dx'} dx'=f(x),
\label{intx}
\end{eqnarray}
we find $f(x)$. In this way, we obtain the  
  topological charge distribution at $\theta=0$.\par
  We should note that this  inversion approach is applicable only when  $x(h)$ is a monotonic function of $h$. Otherwise, the inverted function $h(x)$ is multivalued. We do not treat this case in the present paper. \par
 %%%%%%added%%%
 At $\theta=0$, the function $f(x)$ is usually directly evaluated through
 numerical simulation of the topological charge distribution. By contrast, 
   $f(x)$ is determined in the method proposed by Azcoiti et al. as an implicit function through the process represented by (\ref{zvh}) through (\ref{intx}). 
   Thus, the $x$-$h$ relation is given by (\ref{dfdxQ}). This implies that the functional form of 
   $h(x)$ is obtained  numerically. The relation $h(x)=df(x)/dx$ leads to  $f(x)$ after integration over $x$.\par
 %%%%%%%%%%%%%%2.2
 \subsection{ Qualitative difference between strong and weak coupling behavior}\par
 
 In a previous paper\cite{rf:IKY}, $P(Q)$ of the two-dimensional CP$^{2}$ model is numerically 
 obtained. Now we  study the results from the  point of view
of the  imaginary theta approach.
  In the strong coupling region ($\beta=$ small $\ltsim 1$), $P(Q)$ is 
  approximately given by the Gaussian form
 
       \begin{figure}[htb]
            \parbox{\halftext}{%   %\def\halftext{.471\textwidth}
               \epsfysize= 6 cm
               \hskip.05cm
                 %\epsfbox{Spr1.eps}
                  \epsfbox{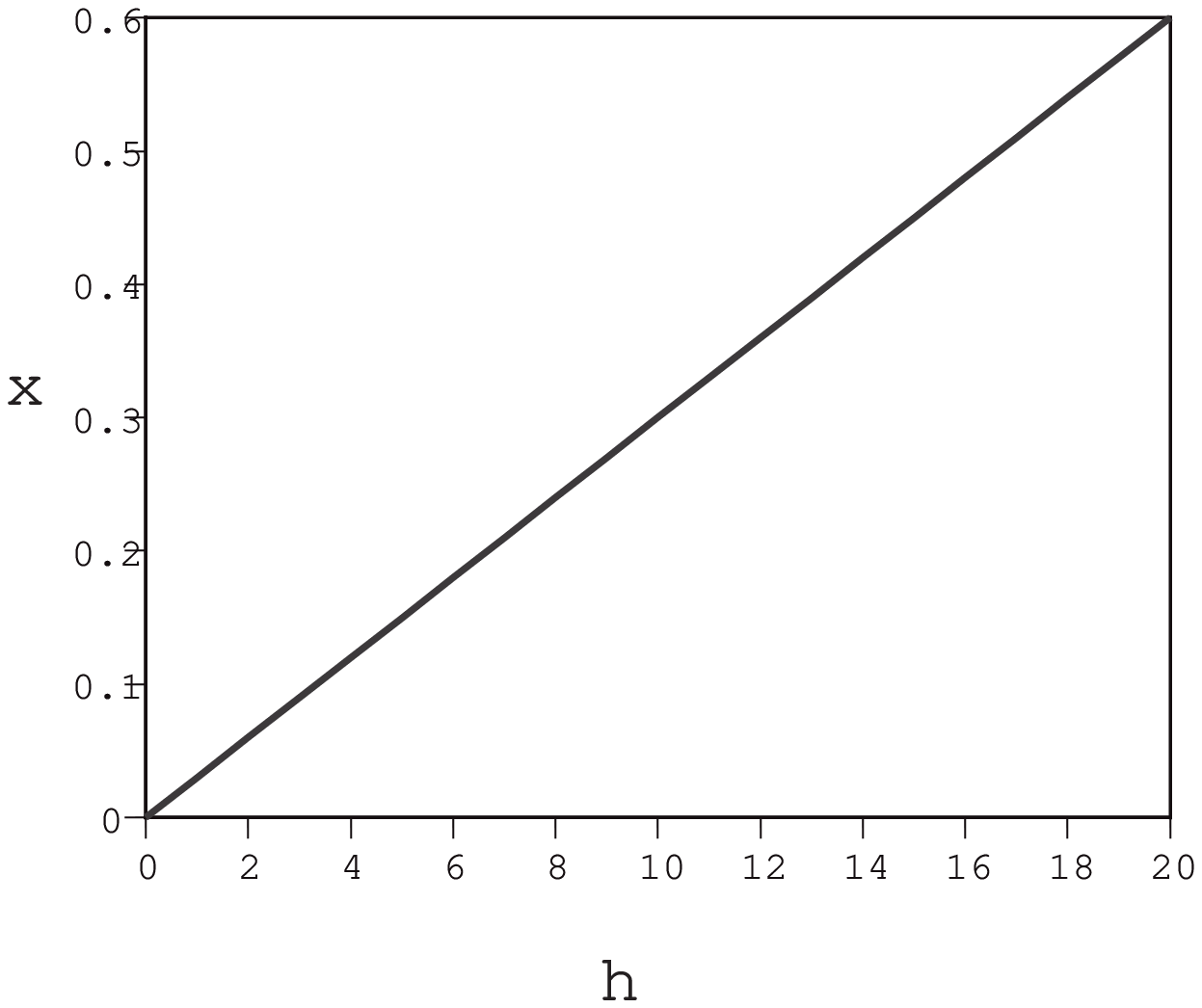}
                     %  \figurebox{6cm}{2cm}
            \label{fig3}
                \caption{Strong coupling case $x$ as a function of $h$.}}
            \hfill
            \parbox{\halftext}{
               \epsfysize= 6 cm
               \hskip.05cm
                 %\epsfbox{Spr1.eps}
                  \epsfbox{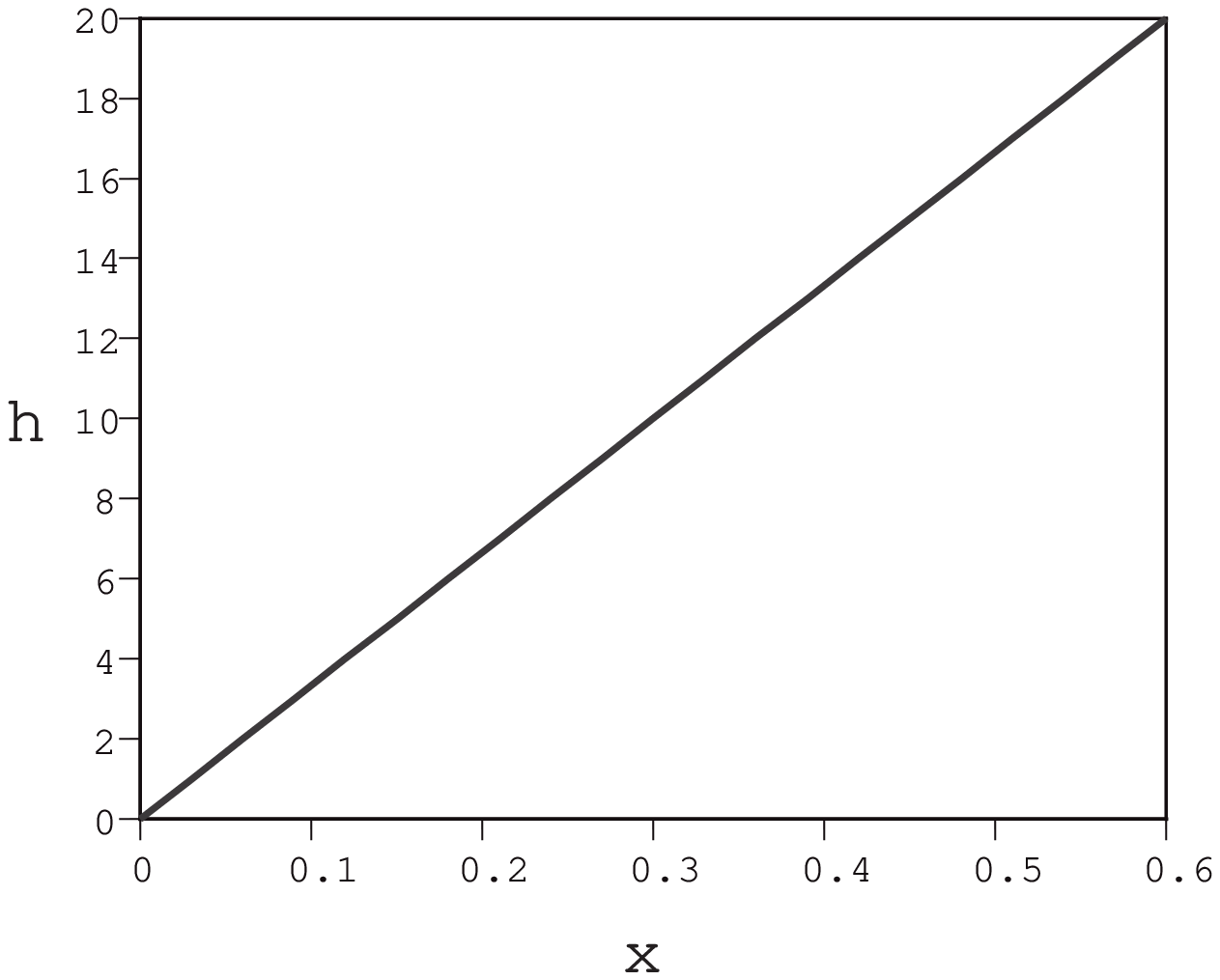}
                     %  \figurebox{6cm}{2cm}
               % \figurebox{6cm}{2cm}
                \caption{Exchange of $x$  and $h$ in Fig. 3.}}
            \label{fig4}
        \end{figure}

\begin{eqnarray}
P(Q)=e^{-\frac{\alpha }{V}Q^2}=e^{-V \alpha x_Q^2},
\label{strong} 
\end{eqnarray}%
and 
\begin{eqnarray}
f(x)=\alpha x^2.
\label{2.24} 
\end{eqnarray}%
In this case, we have  
\begin{eqnarray}
h=f'(x)=2\alpha x.
\label{2.25} 
\end{eqnarray}%
Hence, a linear relation between ($h$ and ${\bar x}$) is expected.\par

In the weak coupling region, the ${\bar x}$-$h$ relation is expected to exhibit ``step-like behavior"(Fig. 5).
In order to make it easy to understand why Fig. 5 is expected, we consider  typical simplified behavior of the $Q$ dependence in the weak coupling region\cite{rf:IKY}, with the  exponent of $P(Q)$ assumed to be  proportional to $|Q|$:

\begin{eqnarray}
P(Q) \sim c^{|Q|} =e^{|Q|\ln c}=e^{-|Q|\ln (1/c)}=e^{-V |x_Q|\ln (1/c)}.
\label{2.26} 
\end{eqnarray}%
The  parameter $c$ is known to be quite small from phenomenological considerations.
 Thus, in this case, we have $f(x)=|x| \ln (1/c)$\cite{rf:IKY}, which we call the ``linear $x$ model" hereafter. Then the relation $h=f'(x)$ is given by
\begin{eqnarray}
h=f'(x)=\left\{ 
\begin{array}{rl}
 \ln(1/c)  &=h_0, \quad x \ge 0, \\
-\ln (1/c) &=-h_0, \quad x < 0.
\end{array}\right .
\label{2.27} 
\end{eqnarray}%
%%%%%%%%%%%%added 1\%\%\%\%\%\%\%\%\%\% added 1\par
The behavior in the $x \ge 0$ region is shown in Fig. 6. Then, interchanging ${\bar x}$ and $h$, we obtain  the $h$-${\bar x}$ relation, shown in Fig. 5. 
From Eq.(\ref{2.27}), Fig.~6 is obtained first. But from the actual numerical simulation, ${\bar x}$ as a function of  the   background $h$ is computed first. Thus, Fig. 5 is put  before Fig. 6.\par 
We set $f_{\rm eff}(x)=f(x)-h x$. By employing the simplified 
functional form ``linear $x$ model", $f(x)=|x| h_0 $, we investigate the three cases (i) $-$ (iii) below.
\begin{itemize}
\item[(i)]
$0<h<h_0$ case:\par
We have

\begin{eqnarray}
f_{\rm eff}(x)=\left\{ 
\begin{array}{l}
-(h_0+h)x, \quad x < 0, \\
(h_0-h)x, \quad x > 0.
\end{array}\right .
\label{2.28} 
\end{eqnarray}
 Then  $\exp(-V f_{\rm eff}(x))$ is peaked at 
$x=0$,    and we expect 
${\bar x}\sim 0$.\par

\item[(ii)] 
$h=h_0>0$ case:\par
In this case, we have
\begin{eqnarray}
f_{\rm eff}(x)=\left\{ 
\begin{array}{r}
 -2h_0 x, \quad x < 0, \\
 \quad \ 0, \quad  x > 0.
\end{array}\right .
\label{2.29} 
\end{eqnarray}
Then $\exp(-V f_{\rm eff}(x))$ is constant in the $x>0$ region, and we expect 
$\bar x$ will take a positive value between 0 and the maximum possible value; that is, $\bar x$ is undetermined.\par

\item[(iii)]
$h>h_0>0$ case:\par
Here we have
\begin{eqnarray}
f_{\rm eff}(x)=\left\{ 
\begin{array}{l}
-(h_0+h)x, \quad x < 0, \\
-(h-h_0)x, \quad x > 0,
\end{array}\right .
\label{2.30} 
\end{eqnarray}
 where $h-h_0$ is positive. Then $\exp(-V f_{\rm eff}(x))$ favors as large a value of  $x$
 as possible and we expect that $\bar x $ is given approximately by the  maximum  possible value. 
 Since $Q$ is bounded from above in the finite volume case, we have $\bar x \ltsim Q_{\rm max}/V$. With periodic boundary conditions, $|Q| \ltsim \frac{V}{2}$ is the limit.\cite{rf:Wiese} Then $|x_Q|$ is bounded from above by $ x_0 \ (\ltsim 1/2)$.  \par
\end{itemize}

%%%%%%%%%%%%added 1\%\%\%\%\%\%\%\%\%\% added 1\par

 %-------------------------------------------------------------------------
      \begin{figure}[htb]
            \parbox{\halftext}{%   %\def\halftext{.471\textwidth}
               \epsfysize= 6 cm
               \hskip.05cm
                 %\epsfbox{Spr1.eps}
                  \epsfbox{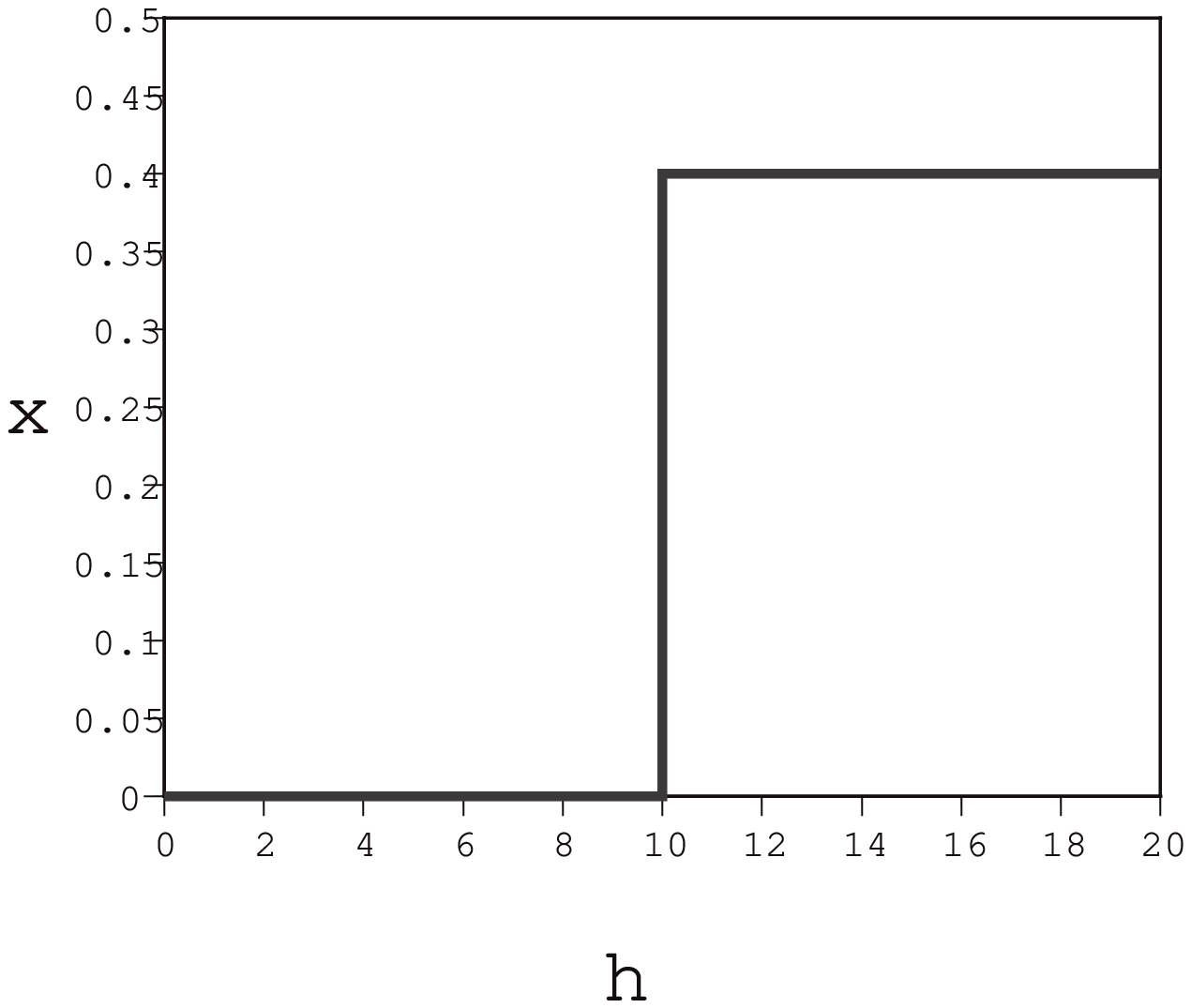}
                     %  \figurebox{6cm}{2cm}
                     \label{fig5}
                \caption{Weak coupling case. The $x$-$h$ relation is inferred from Fig. 6. A step-like increase at $h=h_0$ is expected. The value $h_0$ is tentatively taken to be 10.0 here.}}
            \hfill
            \parbox{\halftext}{
               \epsfysize= 6 cm
               \hskip.05cm
                 %\epsfbox{Spr1.eps}
                  \epsfbox{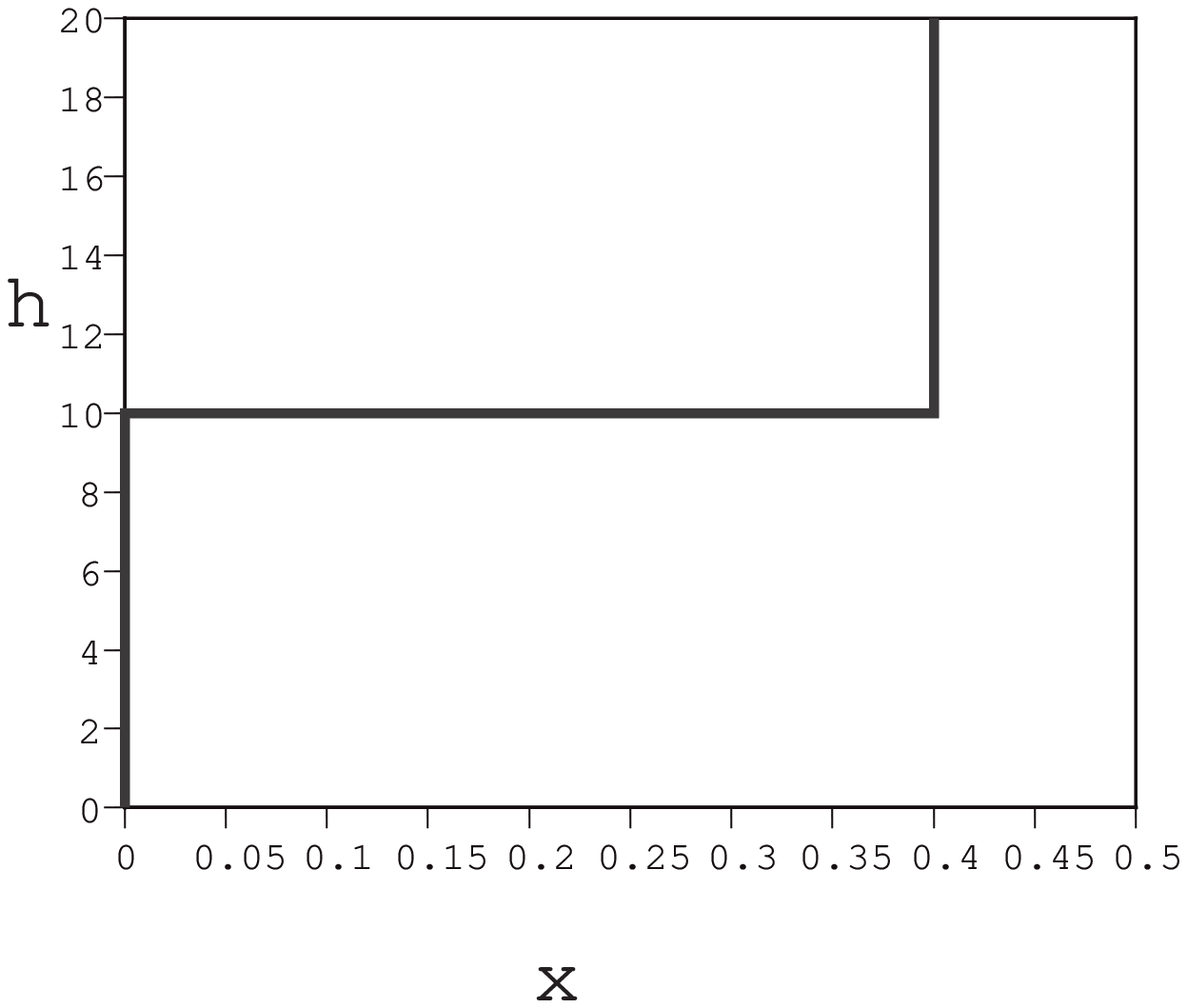}
                     %  \figurebox{6cm}{2cm}
               % \figurebox{6cm}{2cm}
                \caption{A simplified model in which $f(x)=|x| \ln (1/c)$ gives $h=h_0=\ln (1/c)\sim 10.0$ for $x>0$.}}
                 \label{fig6}
        \end{figure}

\vskip 1cm

Summarizing  (i) $-$ (iii), we have found the qualitative behavior of $x$ in the weak coupling region as shown in Fig.5, namely, ``step-like behavior" of $\bar x $ as a function of the background $h$. 

 The expectation value ${\bar x}$ is expected to be quite small in 
 $0 \le h < h_0$ region, while ${\bar x}$  becomes approximately $x_0$ in the $h > h_0$  region. The position of the step-like increase of ${\bar x}$  is expected to be located at 
 $h=h_0=\ln (1/c)$.\par
%%%%%%%
We now give a short comment. In the analysis of finite density QCD, the chemical potential $\m$ and the nucleon density $n\ (=N/V)$ (where $N$ and $V$ denote the nucleon number and the volume of the system) enter in place of the parameter $h$ and the topological charge density $x$. The low temperature, $T \sim 0$, regime of QCD corresponds to the weak coupling,  large $\b$, regime in the CP$^{N-1}$ model, where the topological charge excitation is suppressed due to the fact that only a small fluctuation of the gauge field is allowed. The step-like behavior schematically shown in Fig. 5 is expected to occur in finite density QCD analysis in  $T \sim 0$.\cite{rf:KogSteph}$^{,}$\footnote{ The step-like increase of $n$ at $\m=\m_0$ is schematically given. [See  Figure 8.10(a) in the textbook of Kogut and Stephanov.] See also the paper by S. Kratochvila and Ph. de Forcrand.\cite{rf:Forc1}$^{,}$\cite{rf:Forc2} The results (i) $-$ (iii) of this section are quite similar to  the results presented in Fig. 2 (the $T<T_C$ case) of Ref. \citen{rf:Forc1}. \par}
%%%%%%%

%{rf:KogSteph} 
%-------------------------------------------------------------------------
%   \begin{figure}
% \epsfysize= 4 cm
%\hskip.05cm
%\epsfbox{Spr1.eps}
%\epsfbox{fig2-3b.eps}
%   \caption{}
%   \label{fig2-3b:1}
%   \end{figure}
%

%%%%%%%%%%%%%%%%%%%%%%%%%%%%%%%%%%%%%%%%%%%%%%%%%%%%%%%%%%%
\section{Numerical calculation of the CP$^2$ model}
\label{sec:numerical}
%%%%%%%%%%%%%%%%%%%%%%%%%%%%%%%%%%%%%%%%%%%%%%%%%%%%%%%%%%%%%%%%%%%%%
\subsection{Numerical results}

\begin{itemize}
\item[(1)] As described in \S 2, the expectation value ${\bar  x}(h)   $ was obtained by numerical 
simulation.  The expectation values ${\bar  x}(h)  $ were calculated at various values of the background source
 points $h$.\par
\item[(2)] The saddle point method  gives the relation
\begin{eqnarray}
h=f'({\bar x}).
\label{} 
\end{eqnarray}%
\end {itemize} 
The above (1) and (2) give  the relation between ${\bar x}(h) $ and $h$ at many
 points. From this $x$-$h$ relation, the  form of $f'({\bar x})$ as a function of $\bar x$  is
obtained by fitting  the calculated points. Once the functional form $f'(\xi)$ is 
obtained from  this fitting process, we obtain $f(x)$ itself by integrating   
$f'(\xi)$,

\begin{eqnarray}
f(x)=\int ^x _0 f'(\xi) d\xi.
\label{} 
\end{eqnarray}%
In Ref. \citen{rf:IKY}, we discussed  the ``direct method" and the ``indirect method". The latter is equivalent to the fitting 
method. The imaginary theta method is simply a candidate for the  ``indirect 
method". \par
Now we  present the results of the numerical simulation of the CP$^2$ model using the  imaginary theta method. For each $\b$ and $h$, the number of measurements was set to $10^5$.

%\Fig(3-1)
%-------------------------------------------------------------------------
   \begin{figure}
 \epsfysize= 9 cm
\hskip.05cm
%\epsfbox{Spr1.eps}
\epsfbox{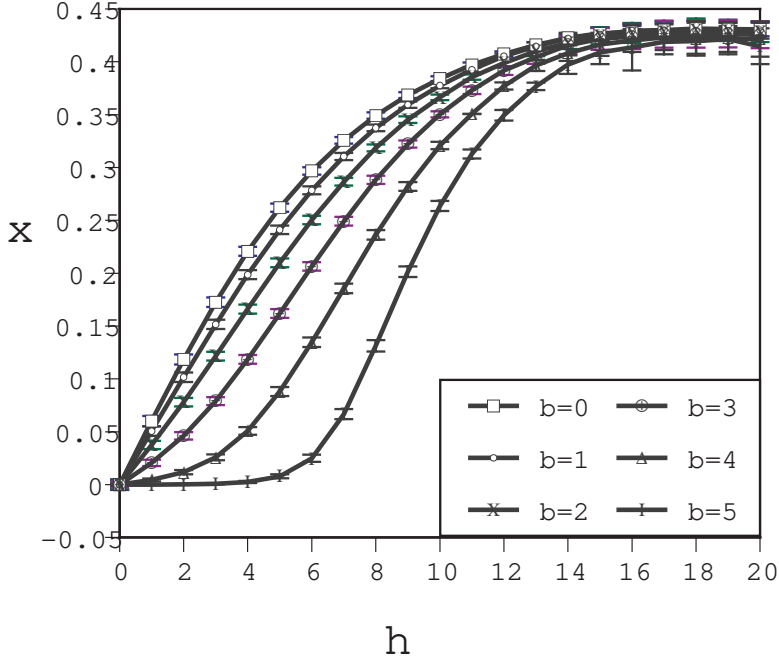}
   \caption{Expectation value $x$ for various values of the background $h$. The values of the inverse coupling $\b$ are 0.0, 1.0, 2.0, 3.0, 4.0 and 5.0. The lattice size is $L=50$ in all cases. }
   \label{fig7}
   \end{figure}

 Figure \ref{fig7} displays the expectation value $\bar x$  with the 
volume $V=L^2 \  (L=50)$ for various values of $h=-{\rm Im} \t$ at 
$\b=0.0, 1.0, 2.0, 3.0, 4.0 $ and $5.0$.  
In the strong coupling cases ($\b=0.0, 1.0)$, $\bar x$ increases linearly as a function of  $h$ from the 
origin: 
\begin{eqnarray}
{\bar x} \sim \frac{1}{2 \a} h. \quad (h=0.0 \sim 5.0)
\label{} 
\end{eqnarray}%

For much larger values of $h$, $\bar x$ exhibits convergence due to the restriction 
that the topological charge cannot exceed $V/2$ in a finite volume\cite{rf:Wiese}. 
Thus ${\bar x}$ is bounded from above:  
\begin{eqnarray}
{\bar  x} < x_0 \ltsim \frac{1}{2}.
\label{} 
\end{eqnarray}%

  \undertext { Step-like behavior } is observed in the weak coupling regions. In 
  weak coupling cases ($\b \gtsim 5.0 $), ${\bar x}$ is strongly suppressed 
  in comparison with the strong coupling cases in the region of small $h$ ($h \ltsim 5.0)$,

\begin{eqnarray}
{\bar x} _{\rm weak} \ll {\bar x}_{\rm strong}.
\label{eq.5} 
\end{eqnarray}%

In regions of larger $h$,  $\bar x$  begins to increase  rapidly and reaches ${\bar x} 
\sim x_0/2 $ at $h \sim h_0 (\b )$. In  $h > h_0 (\b )$ region,  $\bar x$  
 begins to converge to a constant due to the restriction ${\bar x} < x_0. $ \par

The position $ h_0 $ of the step-like increase  depends on the coupling constant; that is, $ h_0 $  is a function of $\b$. The observed step-like behavior is not as sharp as the simple sharp step-like increase mentioned in  \S 2 (Fig. 5),
 but it is clearly observed.  For $\b=5.0$, for example, we have

      \begin{figure}[htb]
            \parbox{\halftext}{%   %\def\halftext{.471\textwidth}
               \epsfysize= 6 cm
               \hskip.05cm
                 %\epsfbox{Spr1.eps}
                  \epsfbox{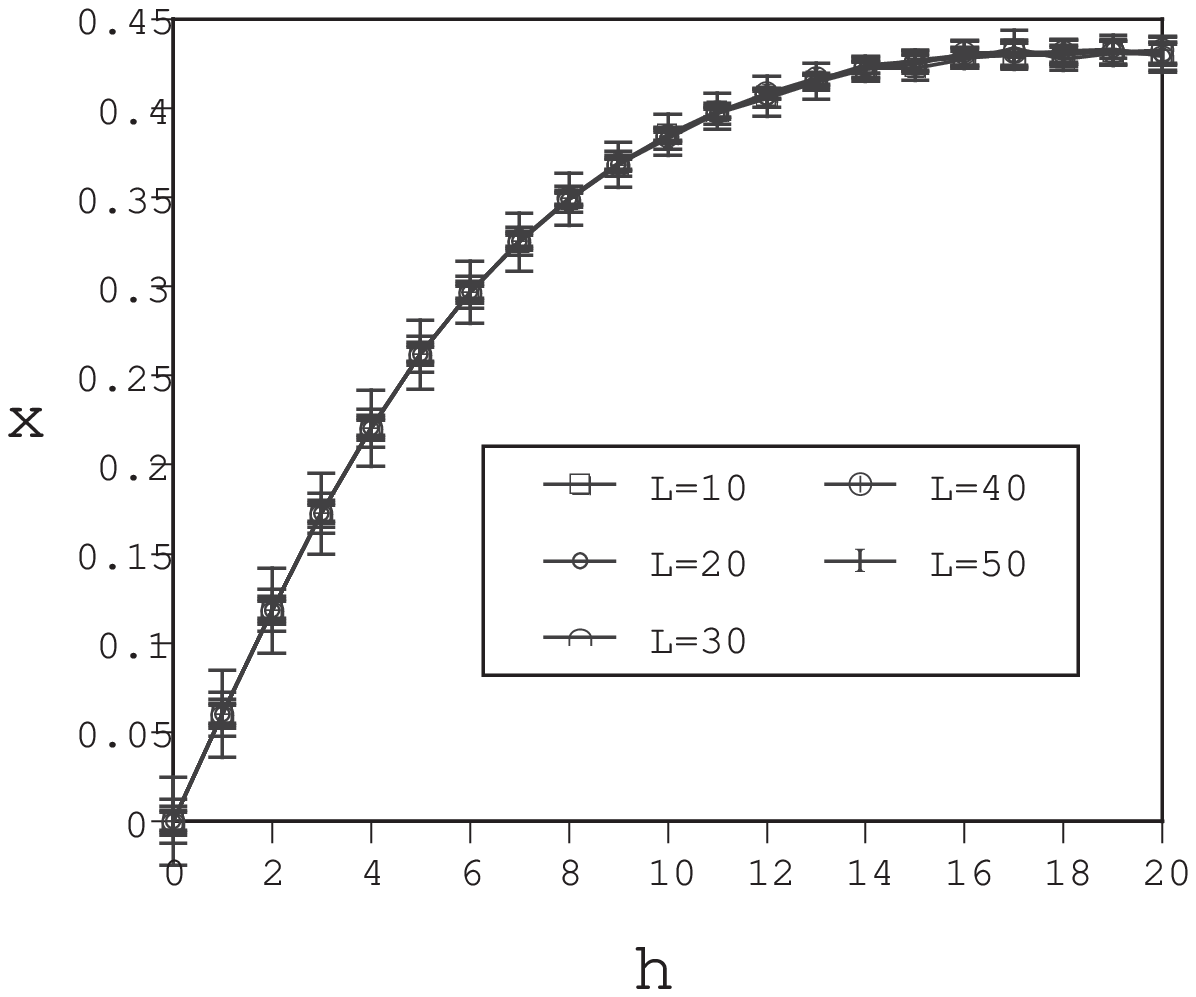}
                     %  \figurebox{6cm}{2cm}
            \label{fig8}
                \caption{Expectation value $x$ in the  strong coupling case ($\b=0.0$). 
   Lattice sizes are $L=$10, 20, 30, 40 and 50. }}
            \hfill
            \parbox{\halftext}{
               \epsfysize= 6 cm
               \hskip.05cm
                 %\epsfbox{Spr1.eps}
                  \epsfbox{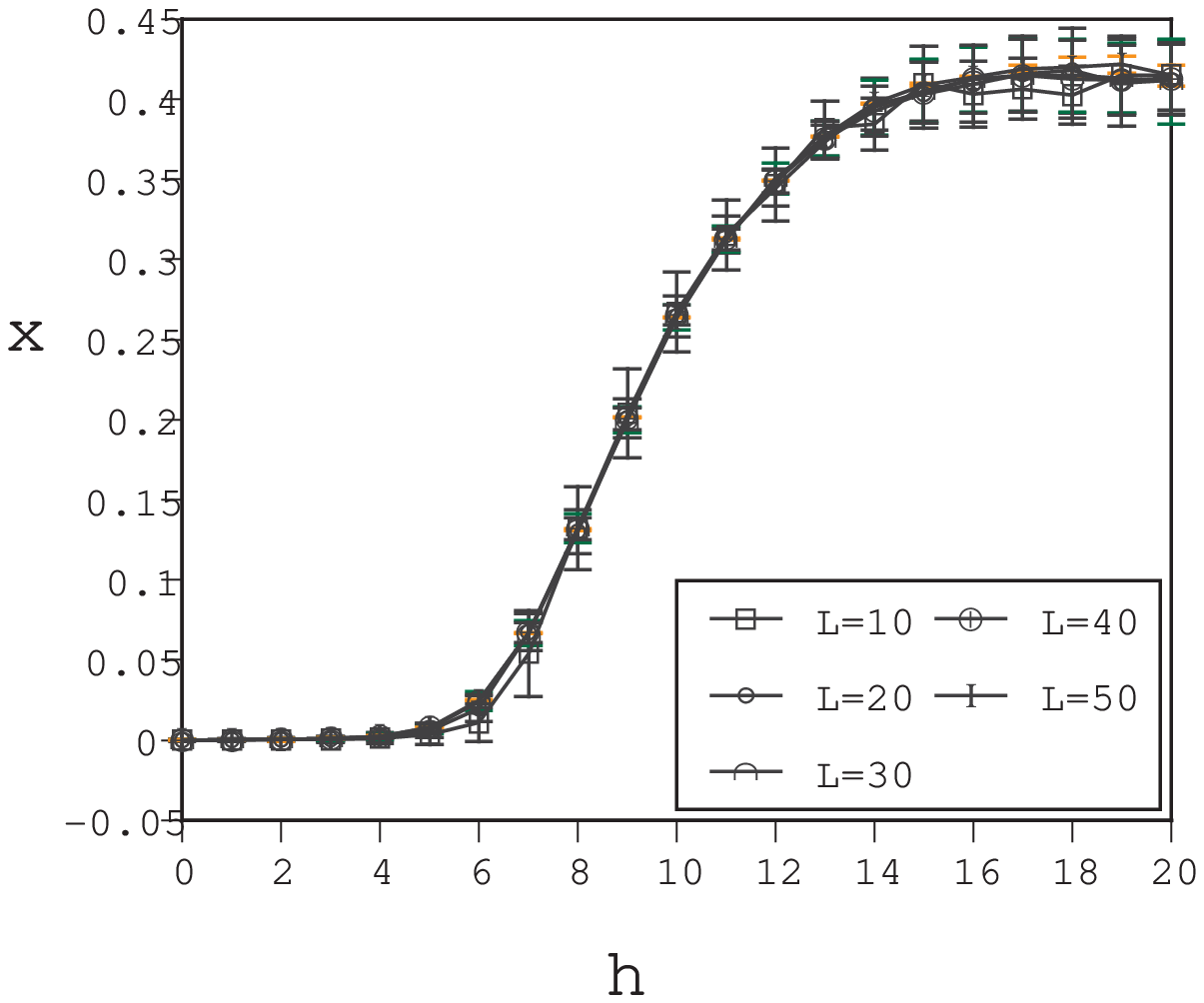}
                     %  \figurebox{6cm}{2cm}
               % \figurebox{6cm}{2cm}
                \caption{Expectation value $x$ for various values of $h$ in the  weak coupling case ($\b=5.0$).   Lattice sizes are $L=$10, 20, 30, 40 and 50.}}
            \label{fig9}
        \end{figure}
\begin{eqnarray}
\begin{array}{rl}
 &  h_0 =\ln (1/c) \sim 10.0. 
\end{array}
\label{} 
\end{eqnarray}%
Note that all $\bar x$ in Fig. 7  are monotonic functions of $h$.\par 
Figure 8 displays $\bar x$ in the strong coupling case ($\b =0$) for various 
sizes, $L=10, 20,30,40$   and $ 50$. The values of  $\bar x$ for different sizes 
coincide and exhibit linear dependence on $h$. Because $h=f'(x)$, a linear dependence 
of $\bar x$ on $h$ for $h \ltsim 5.0$, i.e.,
\begin{eqnarray}
{\bar x} =\frac{1}{2\a} h, 
\label{} 
\end{eqnarray}%
gives
\begin{eqnarray}
f'(x)=2\a x.
\label{} 
\end{eqnarray}%
Integrating this,   we obtain 
\begin{eqnarray}
f(x)= \a x^2,
\label{} 
\end{eqnarray}%
namely a Gaussian distribution for $P(Q)$:
\begin{eqnarray}
P(Q) \propto \exp (-V f(x)) = \exp ( -\frac{\a}{V} Q^2 ).
\label{} 
\end{eqnarray}%

Figure 9 displays  $\bar x$ in the  weak coupling case ($\b=5.0$) for various sizes,  
  $L=10, 20,30,40$   and $ 50$. Smooth step-like behavior is found for  all 
 these sizes. More detailed behavior for small $h$  is shown  in Fig. 10. 
 For small $h$, some dependence  on the value of $L$ is observed.
 The  global behavior for the values  $h=0.0$ -  20.0, however, is almost the same for all values of  $L$ considered here(Fig. 9). \par

       \begin{figure}[htb]
            \parbox{\halftext}{%   %\def\halftext{.471\textwidth}
               \epsfysize= 6 cm
               \hskip.05cm
                 %\epsfbox{Spr1.eps}
                  \epsfbox{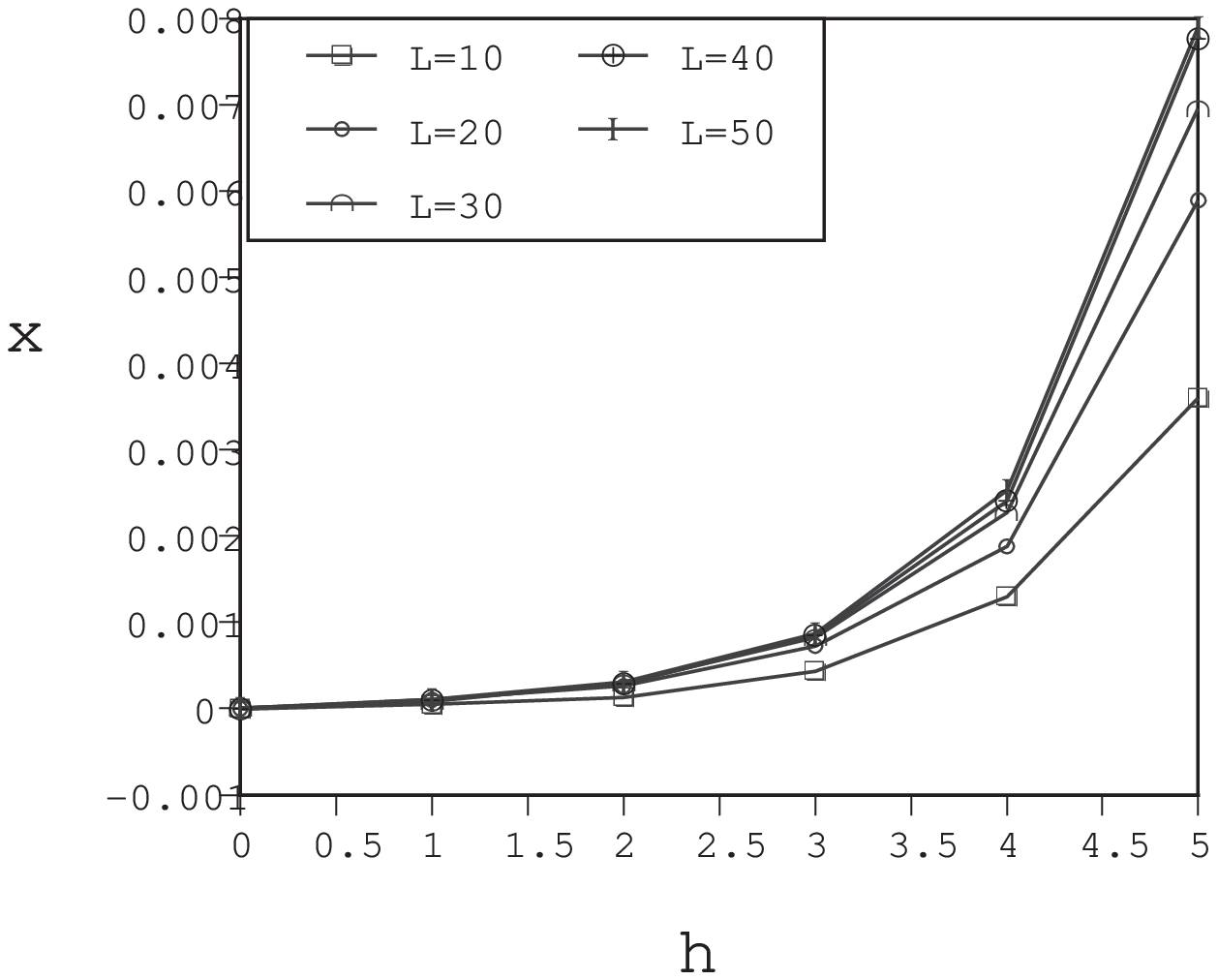}
                     %  \figurebox{6cm}{2cm}
            \label{fig10}
                \caption{Expectation value of $\bar x$ in the  weak coupling case ($\b=5.0$)  at smaller $h \ ( 0.0 \sim 5.0)$. The lattice size dependence is clear for smaller $L \ (10$  - 30). The lattice size dependence is weaker for $L=$ 40 and 50.}}
            \hfill
            \parbox{\halftext}{
               \epsfysize= 6 cm
               \hskip.05cm
                 %\epsfbox{Spr1.eps}
                  \epsfbox{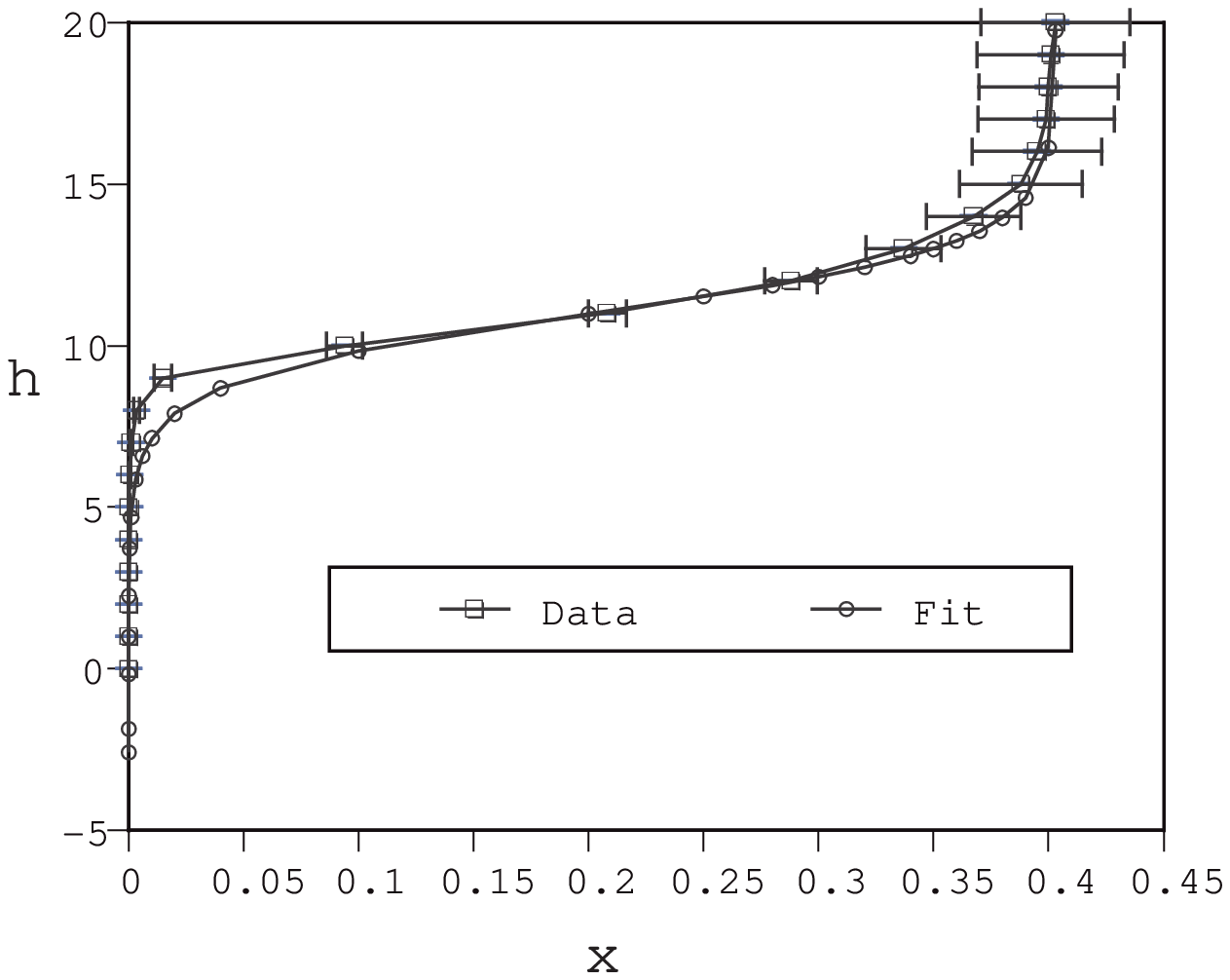}
                     %  \figurebox{6cm}{2cm}
               % \figurebox{6cm}{2cm}
                \caption{Expectation value of $h=f'(x)$ as a function of $x$ for weak coupling ($\b=6.0$) at  $L=$ 50.  
   Both experimental data and a fit [Eq. (\ref{D})] are shown.}}
            \label{fig11}
        \end{figure}

 In Fig. 11,  $h$ vs $\bar x$ in the case of weak coupling, $\b=6.0$, for $L=50$ is  shown. Plateau-like behavior of $h$ at $h=h_0  \sim 11.0$ is clearly seen. It is  smoother than that in the simple case considered in the previous section, but Fig. 11 is the reminiscent of the plateau-like behavior (Fig. 6). To see the detailed behavior for values of $h$ near the origin, a log-log plot is shown in Fig. 12.

\vskip 0.5cm

%%%added2%%%%%%%  analytic continuation %%%%%%\%\%\%\%\%\%\%\%\%\% added 2\par
%%%%%%%%%
\subsection{ Real $\t$, imaginary $\t$ and analytic continuation}\par
Employing a simple ``linear $x$ model", we now investigate the role of the imaginary theta method. The linear $x$ model is defined by
$$ f(x)= |x| h_0,   \qquad h_0=\ln(1/c) >0.$$
Hence, the exponent $f(x)$ of the topological charge distribution is a linear function of the topological charge (divided by the volume). In the case of real $\t$, we have
\begin{eqnarray}
Z_V(\t)=&\sum_{x_n} e^{-V f(x_n)} e^{i V \t x_n}\cr
       =&\sum_{x_n} e^{-V |x_n| h_0} e^{i V \t x_n}.
\label{3.24} 
\end{eqnarray}
In the case of imaginary $\t$, $\t=-ih$ is inserted, and we have
\begin{eqnarray}
Z_V(h) &=&\sum_{x_n} e^{-V |x_n| h_0} e^{ V x_n h }    \qquad       \cr
       &=&({\rm negative } \ Q)+(Q=0)+({\rm positive} \ Q)         \cr
       &=&\sum_{n=1}^{N} e^{-h_0 n-h n}+1+\sum_{n=1}^{N} e^{-h_0 n+h n}\cr
       &=&\sum_{n=1}^{N} t^n +1+\sum_{n=1}^{N} s^n,        
\label{3.25} 
\end{eqnarray}
 where $x_n=n/V=Q/V$, $s=e^{-h_0+h}$ and $t=e^{-h_0-h}$, and the finite volume imposes an upper  bound of the topological charge $Q$ at $N$.\par 
 Then the partition function is given by 
\begin{eqnarray}
Z_V(h) =\frac{t-t^{(N+1)}}{1-t}+1+\frac{s-s^{(N+1)}}{1-s}.
\label{3.26} 
\end{eqnarray}
In the case of real $\t$, $s(\t)=e^{-h_0} e^{i\t}$ and $t(\t)=e^{-h_0} e^{-i\t}$ lead to 
$|s(\t)|<1$ and $|t(\t)|<1$. Thus, $s^{N+1}$ and $t^{N+1}$ approach zero, and Eq. (\ref{3.26}) becomes 
\begin{eqnarray}
Z_V(\t) &\sim &\frac{t(\t)}{1-t(\t)}+1+\frac{s(\t)}{1-s(\t)}\cr
&=&\frac{1-c^2}{1-2c\cos \t +c^2},
\label{A} 
\end{eqnarray}
where $c=e^{-h_0} \ll 1$.\par
In order to address the  question of whether it is possible  to extend real $\t$ to imaginary $\t$, let us study the following two cases.\par

\begin{itemize}
\item[(1)] Imaginary $\t$: $\t=-i h$, with $|h|<h_0$\par
In this case, $s=e^{-h_0+h}<1$  and $t=e^{-h_0-h}<1$, and both $s^{N+1}$ and $t^{N+1}$  can be safely ignored. Then we obtain 
\begin{eqnarray}
Z_V(h) \to \frac{t}{1-t}+1+\frac{s}{1-s}\cr
=\frac{1-c^2}{1-2c\cosh h +c^2}.
\label{B} 
\end{eqnarray}
Equation (\ref{B}) is then simply analytically continued to Eq.(\ref{A}) by taking $ h \to i \t$.\par  

\item[(2)] Imaginary $\t$: $\t=-i h$, with $|h|>h_0$\par
We consider $h>h_0>0.$ It should be noted that

\begin{eqnarray}
t=e^{-h_0-h}<1, \qquad s=e^{-h_0+h}>1.
\label{3.29} 
\end{eqnarray}
In this case, $s$ is greater than unity. Thus, the leading contribution to $Z_V(h)$ is given by 

\begin{eqnarray}
Z_V(h) \sim \frac{-s^{N+1}}{1-s}.
\label{C} 
\end{eqnarray}
From Eq. (\ref{C}), the expectation value of $x$ is 
\begin{eqnarray}
% {\bar x}= \frac{1}{V}\frac{dZ/dh}{Z}\sim \frac{N}{V}={\rm finite}.
 {\bar x}= \frac{1}{V}\frac{dZ/dh}{Z}\sim \frac{N}{V}={\rm finite}=x_0 \ltsim 1/2.    \label{3.31} 
\end{eqnarray}
%
%Important point is that once leading contribution  Eq.(\ref{C}) is taken, 
%non leading  contribution $\left( \frac{t}{1-t}+1+\frac{s}{1-s} \right)$  is
% lost and we can not recover Eq.(\ref{B}) from Eq.(\ref{C}).\par
 An important point is that $Z_V (h)$ is not simply obtained  in the $h > h_0$ region by analytic continuation from Eq. (\ref{A}) by taking $\theta \to -ih$. Rather, $Z_V (h)$ is given by Eq. (\ref{C}),  which is completely different from the analytic continuation form Eq. (\ref{B})  in  the $h > h_0$ region.\par
 \end{itemize}
\vskip 1cm

This simple  ``linear $x$ model" provides the  important lesson that there is 
a case in which imaginary $\t$ does not yield the real $\t$ result by analytic 
continuation. 
Rather, the imaginary $\t$ method  is used as a fitting procedure for the 
topological charge distribution $P(Q)$ at $\t=0$.\cite{rf:ACGL} 
\par
 In the weak coupling region, ${\bar x}$ exhibits step-like behavior as a function of  $h$.  For   $h \ltsim h_0=\ln (1/c)$, ${\bar x}$ is close to zero. Then, it increases suddenly at   $h\sim h_0$ and then satisfies  
 ${\bar x}\sim x_0$ for   $h\gtsim h_0$.  Actually, the $h$-${\bar x}$ relation 
 obtained numerically does not exhibit an abrupt  step-like increase at $h \sim h_0$, but a somewhat gentle 
  one. Simple functional form representing this ``gentle" step-like 
 behavior is expressed    as a function of  $h$ as

\begin{eqnarray}
x=\frac{x_0 e^{c_e (h-h_0)}}{1+e^{c_e (h-h_0)}},
\label{3.32} 
\end{eqnarray}
where ${\bar x}$ is written as $x$. The parameter values are
$$h_0\sim 11.0, \ x_0 \sim 0.4031, \ c_e\sim 0.95$$
for $\b=6.0$ and  $L=50$.\par

      \begin{figure}[htb]
            \parbox{\halftext}{%   %\def\halftext{.471\textwidth}
               \epsfysize= 5.5 cm
               \hskip.05cm
                 %\epsfbox{Spr1.eps}
                  \epsfbox{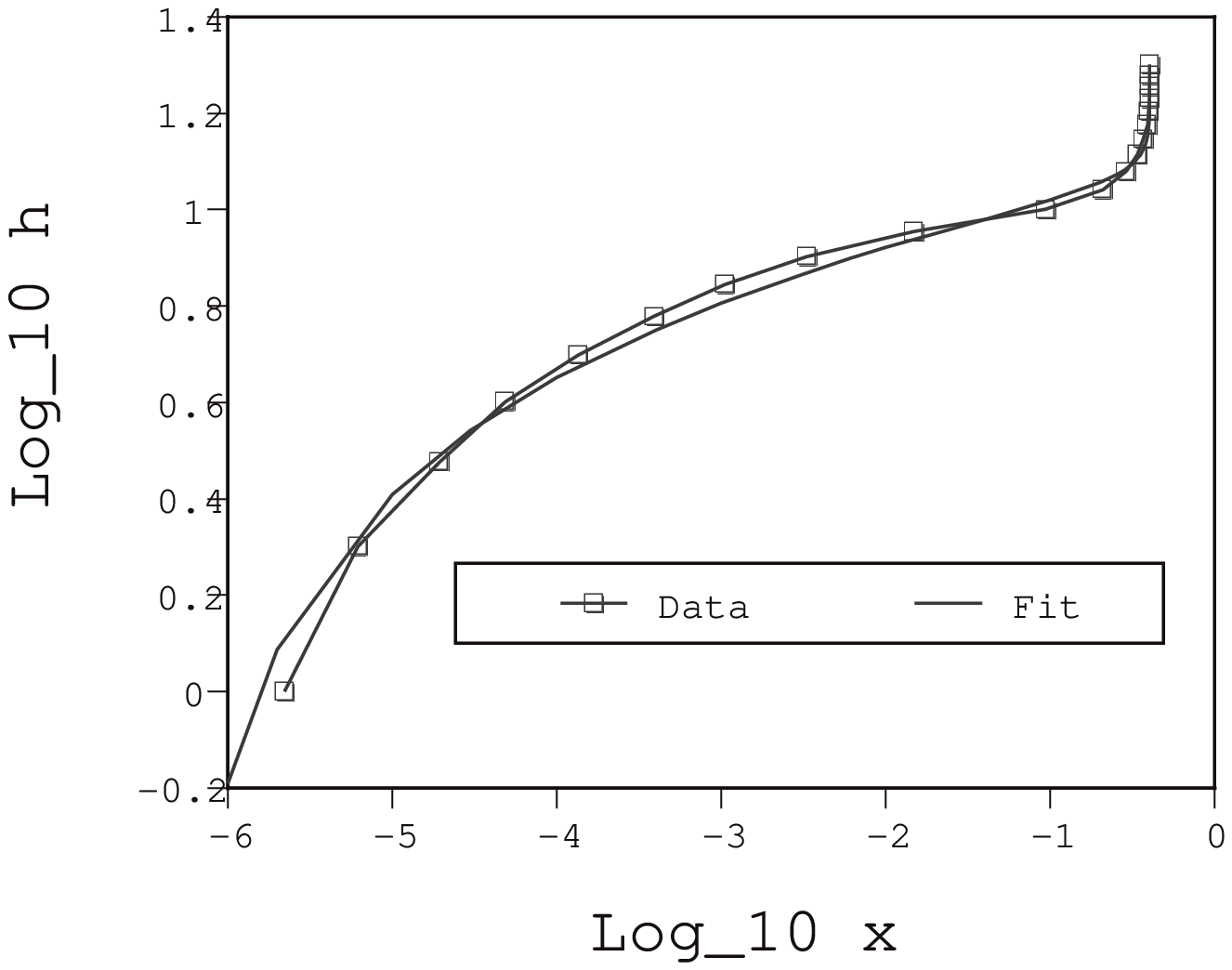}
                     %  \figurebox{6cm}{2cm}
            \label{fig12 }
                \caption{In order to see the detailed behavior of Fig. 11 for  smaller $h$, a log-log plot is shown for the same set of numerical data as in Fig. 11.}}
            \hfill
            \parbox{\halftext}{
               \epsfysize= 5 cm
               \hskip.2cm
                 %\epsfbox{Spr1.eps}
                  \epsfbox{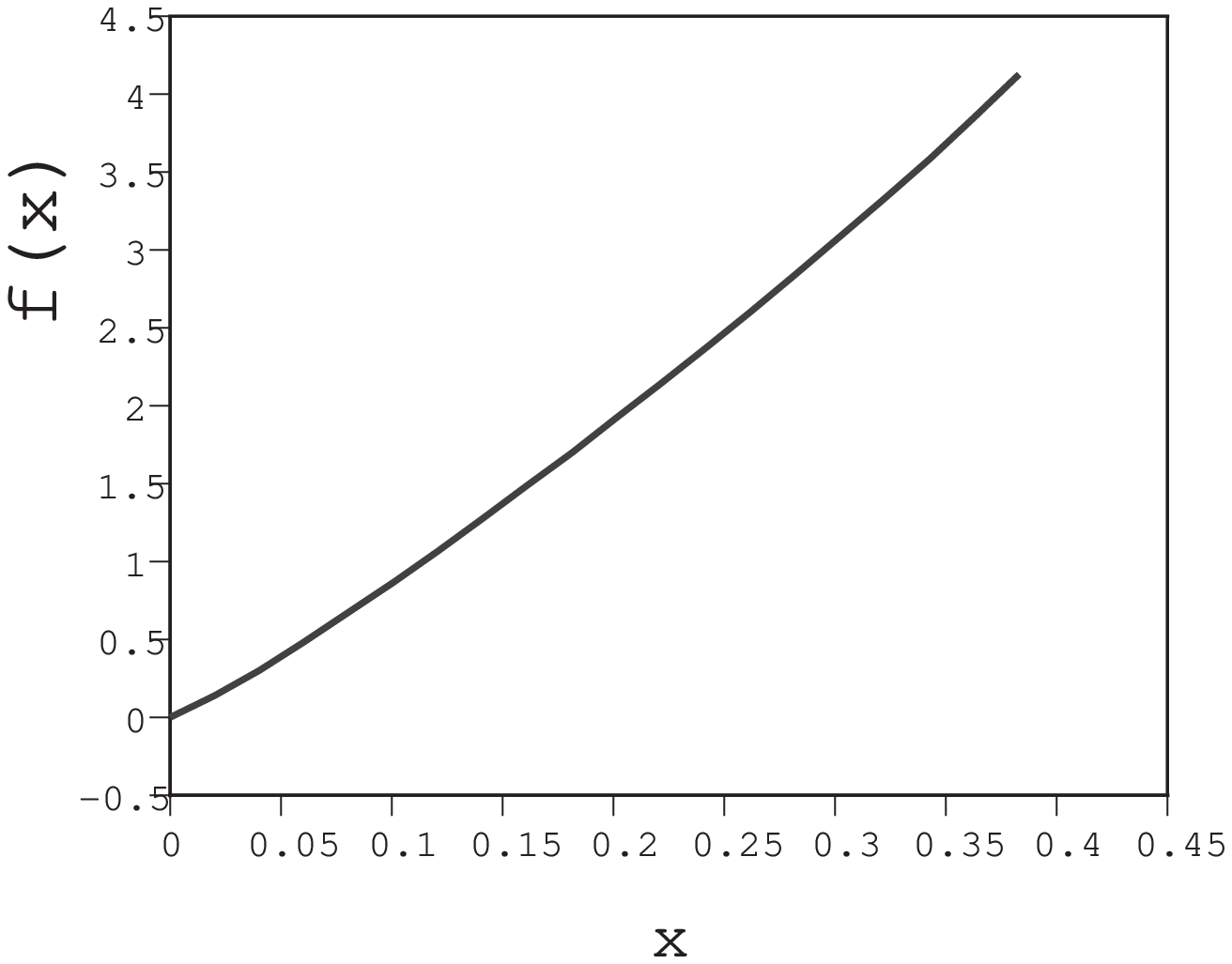}
                     %  \figurebox{6cm}{2cm}
               % \figurebox{6cm}{2cm}
                \caption{Integrating Eq. (\ref{D}), we obtain the analytical form $f(x)$ [Eq. (\ref{E})] for this case.}}
            \label{fig13}
        \end{figure}

This relation is easily inverted. We have 

\begin{eqnarray}
e^{c_e (h-h_0)}=\frac{x}{x_0-x}
\label{3.33} 
\end{eqnarray}
and
\begin{eqnarray}
h=h_0+\frac{1}{c_e} \{\ln x -\ln (x_0-x)  \}.
\label{D} 
\end{eqnarray}
Since $h$ is given by $h=f'(x)$, we integrate Eq. (\ref{D}) and obtain

\begin{eqnarray}
f(x)&=&\int^x f'(x')dx'=\int^x h(x')dx' \qquad \cr
 \qquad  \qquad &=&h_0 x+\frac{1}{c_e} \{x\ln x +(x_0-x)\ln (x_0-x)  \}+d,
\label{E} 
\end{eqnarray}
where $d$ is an integration constant. If we assume $f(0)=0$, $d$ becomes 
\begin{eqnarray}
d=\frac{-1}{c_e}x_0 \ln x_0.
\label{3.36} 
\end{eqnarray}
The value of  $h$ for $\b=6.0$ and $L=50$ is plotted as a function of  $x={\bar x}$ in Fig. 11 and  12. In these figures, the fitting function given in Eq. (\ref{D}) is also shown. The integrated  $f(x)$  is plotted as a function of $x$ in
 Fig. 13.  An almost linear dependence of  $f(x)$ on $x$ is observed in the weak coupling region.  

 %%%%%%%%%%added 2%%%%%%%% \%\%\%\%\%\%\%\%\%\% added 2\par

%%%%%%%%%%%%%%%%%%%%%%%%%%%%%%%%%%%%%%%%%%%%%%%%%%%%%%%%%%%
\section{Conclusions and discussion}
%%%%%%%%%%%%%%%%%%%%%%%%%%%%%%%%%%%%%%%%%%%%%%%%%%%%%%%%%%%%%%%%%%%%%
%\subsection{}
%{\bf  }\par
The inversion approach\cite{rf:ACGL} based on the imaginary theta method\cite{rf:BhDa} was investigated in both strong  and weak coupling regions. 
In the strong coupling region, the expectation value ${\bar x}$ exhibits linear dependence on $h$ for $h \ltsim 5.0$.
In the weak coupling region,  ${\bar x}$ is much smaller than that in the  strong coupling region:
\begin{eqnarray}
{\bar x}({\rm weak})  \ll {\bar x}({\rm strong})  \quad {\rm for}\  h \ltsim 5.0.
\label{} 
\end{eqnarray}
The expectation value ${\bar x}$ is expected to be quite small in the region 
 $0 \le h < h_0$, while ${\bar x}$  is approximately equal to $x_0$ in the region $h > h_0$. The position of the step-like increase of ${\bar x}$  is expected to be located at  $h=h_0=\ln (1/c)$. 
\par

The expectation value ${\bar x}$ thus displays a step-like increase  at $h \sim h_0=\ln(1/c) \sim 10.0 \sim 11.0$. The parameter $c$ represents the probability of a single topological charge excitation in the weak coupling region. The value of $c$ is quite small.\par

 We have numerically calculated ${\bar x}$ for each $h$, thus obtaining ${\bar x}$ as a function of $h$. Then we have obtained $h$ as a function of ${\bar x}$ by inverting ${\bar x}$ and $h$. 
In the large $V$ limit, $h$ is identified with $f'(x)$, and ${\bar x}$ is simply written as $x$.  
If we fit $h$ with an  appropriate functional form $h_{\rm fit}(x)$, then $f(x)$ is obtained by integrating    $h_{\rm fit}(x)$ as
\begin{eqnarray}
f(x)=\int^{x} f'(x')dx'=\int^{x} h_{\rm fit}(x')dx',
\label{4A} 
\end{eqnarray}
where $f(x)$ denotes $f(x)=-V^{-1} \ln P(Q)$. In this way $f(x)$, namely, $P(Q)$ at $\t=0$, is obtained.\par
This process shows that the inversion approach in the imaginary theta method  is not the analytic continuation from $h$ to non-zero theta, but $P(Q)$ at $\t=0$ is obtained as  one of the products of this approach. 
A simplified  fitting function $h_{\rm fit}(x)$ is presented in \S 3, and the result, $f(x)$, is given for that fitting function. The obtained function $f(x)$ [see Eq. (\ref{4A})] is shown in Fig. 13.    \par
%%%%
%\eject
The purpose of the present paper is to clarify the meaning of the inversion approach proposed by Azcoiti et al. For this purpose, we have chosen a simple model, the CP$^2$ model. In our previous analysis\cite{rf:IKY}, it was shown that this model exhibits qualitatively different behavior of $P(Q)$   in the strong and weak coupling regions. This difference  emerges as that in  the ${\bar x}$-$h$ relation of the imaginary theta method. Since we have employed the standard action, and this model is contaminated by dislocations,  precise information about  continuum physics cannot be obtained.  
%%%%%%%add
%The problem of continuum vacuum is posponed as a future study. 
%%%%%%%%%add
Although  a further investigation  of the continuum limit is left for a future study, 
 what we have clarified here, i.e.,   that the  ${\bar x}$-$h$  relation in the weak coupling region exhibits step-like behavior,   should not be altered if a more realistic model were employed. 
 For example, in our previous analysis~\cite{rf:BISY} of the CP$^3$ model with a fixed-point action, it is shown that 
    the topological quantities,  such as  $P(Q)$ and the expectation value $\langle Q \rangle_\theta$,  exhibit   nice  scaling behavior, and that   $P(Q)$ behaves  differently   in  the strong and weak coupling regions. 
 Specifically,  we studied the behavior of the effective  exponent  of $P(Q)$, $\gamma_{\rm eff}$,  and we found  that $\gamma_{\rm eff}$  in the weak coupling region   is  smaller  than  that in the strong coupling region (Gaussian).   This  suggests the possibility of  step-like behavior of ${\bar x}$-$h$ in the continuum limit. Further study of this point is needed. \par
% From this  ${\bar x}$-$h$ relation,  $f(x)$, namely topological charge distribution $P(Q)$ at $\t=0$, is obtained. \par
  
 ``The step-like increase" of ${\bar x}$ at $h\sim h_0$ in the weak coupling region ($1/\b \sim 0$) is  
 shematically shown in Fig. 5. The results of the actual numerical simulation are shown in Fig. 9. As stated at  the end of \S 2, the $h$-$\bar x $ relation is quite similar to the $\m$-$ n$ (chemical potential vs nucleon density) relation in QCD. For $T\sim 0$ (where $T$ denotes temperature), a step-like increase at  $\m\sim \m_0$ is expected, as shown in Figure 8-10(a) of the textbook of Kogut and Stephanov.\cite{rf:KogSteph} Further investigation of the correspondence between $h$-$\bar x $ in the CP$^{N-1}$ and $\m$-$ n$ relation in QCD is an interesting problem.\par

\par\par\par\par\par\par\par\par\par\par\par\par\par\par\par\par\par\par\par\par\par\par\par\par\par\par\par

\section*{Acknowledgements}
We would like to thank  members of the
 particle physics group of  Yamagata University and Niigata University for valuable discussion at the annual inter-university workshop ``Niigata-Yamagata Gasshuku". A preliminary version of this work was presented there in Nov. 2003.
This work is supported in part by Grants-in-Aid for Scientific Research (C)(2) from the Japan Society for Promotion of Science (No. 15540249) and from the Ministry of Education Science, Sports and Culture (No. 13135213 and No. 13135217). Niigata-Yamagata Gasshuku is financially supported by YITP, Kyoto University, No. YITP-S-05-02.\par
After the completion of the present manuscript, we received an interesting  mail from Prof. de Forcrand. We thank him for bringing our attention to Refs. \citen{rf:Forc1} and \citen{rf:Forc2}, where we found a close correspondence between $h$-$x$ of the CP$^{N-1}$ model and $\mu$-$B$ (where $B$ denotes the baryon number) in QCD.\par

 %We would like to thank ...........

%\appendix
%\section{First Appendix} %Empty argument \section{} yields `Appendix'. 
%
%\section{Second Appendix}

\end{document}